\documentclass[10pt]{article}
\usepackage{latexsym,graphicx}
\newcommand{\be}{\begin{equation}}
\newcommand{\ee}{\end{equation}}
\def\n{\noindent}
\catcode `\@=11
\catcode `\@=12
\begin{document}
\begin{center}
\large{\bf {Higher Dimensional Cosmological Implications Of A Decay Law For $\Lambda$ Term : 
Expressions For Some Observable Quantities}} \\
\vspace*{10mm}
\normalsize{Anirudh Pradhan
\footnote{Department of Mathematics, Hindu Post-graduate College, 
Zamania-232 331, Ghazipur, India. E-mail: apradhan@iucaa.ernet.in, acpradhan@yahoo.com}
, G. S. Khadekar
\footnote{Department of Mathematics, Nagpur University, Mahatma Jyotiba Phule
Educational Campus, Amravati Road, Nagpur - 440 033, India. E-mail : gkhadekar@yahoo.com, 
gkhadekar@rediffmail.com} 
and Deepak Srivastava 
\footnote{Department of Mathematics, Hindu Post-graduate College, 
Zamania-232 331, Ghazipur, India.}}
\end{center}
\vspace{10mm}
\begin{abstract} 
Implications of cosmological model
with a cosmological term of the form $\Lambda = \beta \frac{\ddot
{a}}{a}$, where $\beta$ is a constant, are analyzed in
multidimensional space time. The proper distance, the luminosity
distance-redshift, the angular diameter distance-redshift, and
look back time-redshift for the model are presented. It has been
shown that such models are found to be compatible with the recent
observations. This work has thus generalized to higher dimensions
the well-know result in four dimensional space time. It is found
that there may be significant difference in principle at least,
from the analogous situation in four dimensional space time.  
\end{abstract}
\smallskip
\n PACS: 98.80.-k, 98.80.Es\\
\n Key words : Cosmology, higher dimensional space time, cosmological constant, 
cosmological tests\\
\newpage
\section{Introduction}
\vspace*{-0.5pt} Higher dimensional space time is an active
research in its attempt to unify gravity with other forces in
nature. This idea is particularly important in the filed of
cosmology since one knows that our universe has much smaller in
its early stage than it today. Indeed the present four dimensional
stage of the universe could have been preceded by higher
dimensional stage, which at later times becomes effectively four
dimensional in the sense that the extra dimension become
un-observably small due to dynamical contraction (Chodos and
Detweiler \cite{ref1}). In this work we consider multidimensional
Robertson Walker (RW) model as a test case. In RW type of
homogenous cosmological model, the dimensionality has a marked
effect on the time temperature relation of the universe and our
universe appears to cool more slowly in higher dimensional space
time (Chatterjee \cite{ref2}). \\

One of the most important and outstanding problems in cosmology is the
cosmological constant problem. The recent observations indicate that
$\Lambda \sim 10^{-55}cm^{-2}$ while particle physics prediction
for $\Lambda$ is greater than this value by a factor of order $10^{120}$.
This discrepancy is known as cosmological constant problem. Some of the recent
discussions on the cosmological constant ``problem'' and
consequence on cosmology with a time-varying cosmological constant
are investigated by Ratra and Peebles \cite{ref3}, Dolgov \cite{ref4} $-$ \cite{ref6}, 
Sahni and Starobinsky \cite{ref7}, Padmanabhan \cite{ref8} 
and Peebles \cite{ref9}. Recent observations by Perlmutter {\it et
al.} \cite{ref10} and Riess {\it et al.} \cite{ref11} strongly favour
a significant and positive value of $\Lambda$. Their finding arise
from the study of more than $50$ type Ia supernovae with redshifts
in the range $0.10 \leq z \leq 0.83$ and these suggest Friedmann
models with negative pressure matter such as a cosmological
constant $(\Lambda)$, domain walls or cosmic strings (Vilenkin \cite{ref12}, 
Garnavich {\it et al.} \cite{ref13}). Recently, Carmeli
and Kuzmenko \cite{ref14} have shown that the cosmological
relativistic theory (Behar and Carmeli \cite{ref15}) predicts the
value for cosmological constant $\Lambda = 1.934\times 10^{-35}
s^{-2}$. This value of ``$\Lambda$'' is in excellent agreement
with the measurements recently obtained by the High-z Supernova
Team and Supernova Cosmological Project (Garnavich {\it et al.} \cite{ref13}; 
Perlmutter {\it et al.} \cite{ref10}; Riess {\it et al.} \cite{ref11}; 
Schmidt {\it et al.} \cite{ref16}). The main conclusion of these observations is 
that the expansion of the universe is accelerating.\\

Several ans$\ddot{a}$tz have been proposed in which the $\Lambda$
term decays with time (see Refs. Gasperini \cite{ref17},
Gasperini \cite{ref18}, Berman \cite{ref19}, Freese {\it et
al.} \cite{ref34}, $\ddot{O}$zer and Taha \cite{ref20}, Peebles and
Ratra \cite{ref21}, Chen and Hu \cite{ref22}, Abdussattar and
Viswakarma \cite{ref23}, Gariel and Le Denmat \cite{ref24}, Pradhan
{\it et al.} \cite{ref25}). Of the special interest is the
ans$\ddot{a}$tz $\Lambda \propto a^{-2}$ (where $a$ is the scale
factor of the Robertson-Walker metric) by Chen and Wu \cite{ref22},
which has been considered/modified by several authors
(Abdel-Rahaman \cite{ref26}, Carvalho {\it et al.} \cite{ref27},
Waga \cite{ref28}, Silveira and Waga \cite{ref29},
Vishwakarma \cite{ref30}). However, not all vacuum decaying
cosmological models predict acceleration. Al-Rawaf and Taha and
Al-Rawaf \cite{ref31} and Overdin and Cooperstock \cite{ref32}
proposed a cosmological model with a cosmological constant of the
form $\Lambda = \beta \frac{\ddot{a}}{a}$, where $a$ is the scale
factor of the universe and $\beta$ is a constant. Following the
same decay law recently Arbab \cite{ref33} have investigated
cosmic acceleration with positive cosmological constant and also
analyze the implication of a model built-in cosmological constant for
four-dimensional space time. The cosmological consequences of this decay 
law are very attractive. This law provides reasonable solutions to the
cosmological puzzles presently known. One of the motivations for
introducing $\Lambda$ term is to reconcile
the age parameter and the density parameter of the universe with recent 
observational data.\\

In this paper by considering cosmological implication of decay law
for $\Lambda$ that proportional to $\frac{\ddot{a}}{a}$, we
discuss the cosmological tests pertaining proper distance,
luminosity distance, angular diameter distance, and look-back time
in the framework of higher dimensional space time  and shown that
Freese {\it et al.} \cite{ref34} model is retrieved from our model
for a particular choice of $A_{0}$ and $ n= 2$. The Einstein-de
Sitter (ES) results are also obtained from our results for the case 
$A_{0} = \frac{1}{2}$ and $n=2$.

\section{The Metric and Field  Equations}
Consider the $(n+2)$-dimensional homogeneous and isotropic model
of the universe represented by the space time
\begin{equation}
\label{eq1} ds^{2} = dt^2 - a^{2}(t)\left[\frac{dr^2}{1-k r^2}+
r^2 dX_{n}^{2}\right],
\end{equation}
where $a(t)$ is the scale factor, $k=0,\; \pm 1$ is the curvature
parameter and $$ dX_{n}^2= d\theta_{1}^2 + sin^2
\theta_{1}d\theta_{2}^2 +...+sin^2 \theta_{1}sin^2
\theta_{2}...sin^2 \theta_{n-1} d\theta_{n}^{2}$$. 
The usual energy-momentum tensor is modified by addition of a term
\begin{equation}
\label{eq2} 
T^{vac}_{ij} = - \Lambda(t) g_{ij},
\end{equation}
where $\Lambda(t)$ is the cosmological term and $g_{ij}$ is the
metric tensor. \\

For the perfect fluid distribution Einstein field equations with
the cosmological constant $\Lambda$ and gravitational constant
$G=1$ may be written as
\begin{equation}
\label{eq3}
 R_{ij} - \frac{1}{2} R g_{ij} = - 8 \pi
T_{ij}-\Lambda(t) g_{ij}.
\end{equation}
The energy-momentum tensor is $T_{ij}$ is defined as
\begin{equation}
\label{eq4} 
T_{ij} = (p + \rho)u_{i}u_{j} - p g_{ij},
\end{equation}
where $p$ and $\rho$ are, respectively, the energy and pressure of
the cosmic fluid, and $u_{i}$ is the fluid four-velocity such that
$u^{i}u_{i} = 1$. The Einstein filed Eqs. (3) and (4) for the
metric (1) take the form
\begin{equation}
\label{eq5} 
\frac{n(n+1)}{2} \left[\frac{\dot{a}^{2}} {a^{2}}+
\frac{k}{a^2}\right]=8 \pi\rho + \Lambda(t),
\end{equation}
\begin{equation}
\label{eq6} 
\frac{n\ddot{a}}{a} + \frac{n(n-1)}{2}\left[\frac{\dot{a}^{2}}
{a^{2}}+\frac{k}{a^2}\right] = - 8 \pi p + \Lambda(t).
\end{equation}
An over dot indicates a derivative with respect to time $t$. The
energy conservation equation $T^{i}_{j;i} = 0$ leads to
\begin{equation}
\label{eq7} \dot{\rho} + (n+1)(\rho + p)H = -\frac{\dot\Lambda}{8 \pi} \; ,
\end{equation}
where $H=\frac{\dot{a}}{a}$ is the Hubble parameter.\\

For complete determinacy of the system, we consider a perfect-gas
equation of state
\begin{equation}
\label{eq8} p = \gamma \rho, ~ ~ 0 \leq \gamma \leq 1.
\end{equation}
It is worth noting here that our approach suffers from a lack of
Lagrangian approach. There is no known way to present a consistent
Lagrangian model satisfying the necessary conditions discussed
in the paper. 

\section{Solution of the Field Equations}
In case of the stiff fluid i.e. $\gamma =1$ in Eq. (8), Equations (5) and (6) with $k=0$ reduces to
\begin{equation}
\label{eq9} 
\frac{\ddot{a}}{a} + n \left(\frac{\dot{a}}{a}\right)^{2} = \frac{2\Lambda(t)}{n}.
\end{equation}
We propose a phenomenological decay law for $\Lambda$ of the form \cite{ref27,ref28}
\begin{equation}
\label{eq10} \Lambda = \beta \left(\frac{\ddot{a}}{a}\right),
\end{equation}
where $\beta$ is constant. Overdin and Cooperstock have pointed out that the model with 
$\Lambda \propto H^{2}$ is equivalent to above form. \\

Using Eq. (10) in Eq. (9) and by integrating we obtain
\begin{equation}
\label{eq11} 
a(t) = \left[\frac{K}{A_{0}} ~ t \right]^{A_{0}},
\end{equation}
where $K$ is an integrating constant and the constant $A_{0}$ has the value
\begin{equation}
\label{eq12} 
A_{0} = \frac{(2 \beta -n)}{2\beta -n(n+1)}.
\end{equation}
By using Eq. (11) in the field equations Eqs. (5) and (6) we obtain
\begin{equation}
\label{eq13} \Lambda(t) =  \frac{n^2\beta (2\beta-n)}{[2\beta
-n(n+1)]^2} \frac{1}{t^{2}} \; ,\;\; \;\;\; \beta \ne \frac{n(n+1)}{2}.
\end{equation}
\begin{equation}
\label{eq14}  \rho(t) =\frac{n(2\beta-n)}{16\pi[2\beta -n(n+1)]}
\frac{1}{t^{2}} \; , \;\;\;\;\; \beta \ne \frac{n(n+1)}{2}.
\end{equation}
The deceleration parameter $q$ is defined as
\begin{equation}
\label{eq15} 
q = -\frac{\ddot{a}a}{\dot{a}^{2}} =\frac{(1-A_{0})}{A_{0}}=\frac{n^2}{(n-2\beta)} \; ,\;\;\;\;\;\beta
\ne \frac{n}{2}
\end{equation}
The density parameter of the universe $\Omega_{m}$ is given by
\begin{equation}
\label{eq16} \Omega_{m} = \frac{16 \pi
\rho}{n(n+1)H^2}=\frac{2\beta
-n(n+1)}{(n+1)(2\beta-n)} \; , \;\;\;\;\;\beta \ne \frac{n}{2},  \; \; \; n \geq 2.
\end{equation}
The density parameter due to vacuum contribution is defined as
$\Omega_{\Lambda}=\frac{2\Lambda}{n(n+1)H^2}.$ Employing Eq. (13),
this gives
\begin{equation}
\label{eq17}
\Omega_{\Lambda}=\frac{2n\beta}{(n+1)(2\beta-n)}\;\;,\;\;\;\;\;\beta
\ne \frac{n}{2}, \; \; \; n \geq 2. 
\end{equation}
From Eqs. (16) and (17), we obtain
 $$\Omega_{m} + \Omega_{\Lambda} = 1.$$
According to high redshift supernovae and CMB, the preliminary
results from the advancing field of cosmology suggest that the
universe may be accelerating universe with a dominant contribution
to its energy density coming in the form of cosmological
$\Lambda$-term. The results, when combined with CMB anisotropy
observations on intermediate angular scales, strongly support a
flat universe
\begin{equation}
\label{eq18} \Omega_{m} + \Omega_{\Lambda} = 1.
\end{equation}
The age of the universe is calculated as
\begin{equation}
\label{eq19} t_{0} = H_{0}^{-1}A_{0}.
\end{equation}

\section{Neoclassical Tests (Proper Distance $d(z)$)}
A photon emitted by a source with coordinate $r=r_{1}$ and
$t=t_{1}$ and received at a time $t_{0}$ by an observer located at
$r=0$. The emitted radiation will follow null geodesics on which
$(\theta_{1},\theta_{2},...\theta_{n})$ are constant. \\

The proper distance between the source and observer is given by
\begin{equation}
\label{eq20}
 d(z)=a_{0}\int_{a}^{a_{0}}\frac{da}{a \dot{a}}\;\;\:,
\end{equation}
$$r_{1}=\int_{t_{1}}^{t_{0}}\frac{dt}{a}=\frac{a_{0}^{-1}H_{0}^{-1}A_{0}}{(1-A_{0})}
\left[ 1- (1+z)^{\frac{A_{0}-1}{A_{0}}}\right].$$ Hence
\begin{equation}
\label{eq21}
 d(z)=r_{1}a_{0}=H_{0}^{-1}\left(\frac{A_{0}}{1-A_{0}}\right)
 \left[ 1- (1+z)^{\frac{A_{0}-1}{A_{0}}}\right],
\end{equation}
where $(1 + z) = \frac{R_{0}}{R}$ = redshift and $a_{0}$ is the present scale factor of the
universe.\\

For small $z$ Eq. (21) reduces to
 \begin{equation}
 \label{eq22}
 H_{0} d(z)= z - \frac{1}{2}A_{0}z^{2} + ...
\end{equation}
By using Eq. (15)
\begin{equation}
\label{eq23}
 H_{0} d(z)= z - \frac{1}{2}(1 + q)z^{2} + ...
\end{equation}
From Eq. (21), it is observed that the distance $d$ is maximum at
$z = \infty$. Hence
\begin{equation}
\label{eq24}
 d(z = \infty) = H_{0}^{-1}\left(\frac{A_{0}}{1-A_{0}}\right)=\frac{H_{0}^{-1}}{n^2}(n-2\beta)
 \end{equation}
Eq. (21) gives the Freese {\it et al.} results for the proper
distance if we choose $n=2$ and  
$$\frac{A_{0}}{(1 - A_{0})} = \frac{2}{(3 \Omega_{m} - 2)}$$
and $d(z)$ is maximum for $\Omega_{m} \to 0 $ (de-Sitter
universe) and minimum for $\Omega_{m} \to 1 $ ES.

\section{Luminosity Distance}
Luminosity distance is the another important concept of
theoretical cosmology of a light source. The luminosity distance is a way of expanding
the amount of light received from a distant object. It is the distance that the object 
appears to have, assuming the inverse square law for the reduction of light intensity
with distance holds. The luminosity distance is {\it not} the actual distance to the
object, because in the real universe the inverse square law does not hold. It is broken
both because the geometry of the universe need not be flat, and because the universe is
expanding. In other words, it is defined in such a way as generalizes the inverse-square 
law of the brightness in the static Euclidean space to an expanding curved space \cite{ref28}.\\

If $d_{L}$ is the luminosity distance to the object, then
\begin{equation}
\label{eq25} 
d_{L} = \left(\frac{L}{4\pi l}\right)^{\frac{1}{2}},
\end{equation}
where $L$ is the total energy emitted by the source per unit time,
$l$ is the apparent luminosity of the object. Therefore one can
write
\begin{equation}
\label{eq26} 
d_{L} = r_{1}a_{0}= d(1 + z).
\end{equation}
Using Eq. (21) equation  (26) reduces to
\begin{equation}
\label{eq27}
 H_{0} d_{L} = (1 + z)\left(\frac{A_{0}}{1-A_{0}}\right)\Big[1 - (1 +
z)^{\frac{A_{0}-1}{A_{0}}}\Big].
\end{equation}
For small value of  $z$, Eq. (27) gives
\begin{equation}
\label{eq28} 
H_{0} d_{L} = z + \frac{1}{2}(1 - q)z^{2} + ...
\end{equation}
or by using Eq. (16)
\begin{equation}
\label{eq29} 
H_{0} d_{L} = z + \Big[1-\left(\frac{n+1}{2}\right)\Omega_{m}\Big]z^{2} +...
\end{equation}
The luminosity distance depends on the cosmological model we have under discussion, and hence
can be used to tell us which cosmological model describe our universe. Unfortunately, however,
the observable quantity is the radiation flux density received from an object, and this can 
only be translated into a luminosity distance if the absolute luminosity of the object is known.
There is no distant astronomical objects for which this is the cases. This problem can however be 
circumvented if there are a population of objects at different distances which are believed to have
the same luminosity; even if that luminosity is not known, it will appear merely as an overall
scaling factor. \\

Such a population object is Type Ia supernovae. These are believed to be caused by the core 
collapse of white dwarf stars when they accrete material to take them over the Chandrasekhar
limit. Accordingly, the progenitors of such supernovae are expected to very similar, leading
to supernovae of a characteristic brightness. This already gives good standard candle, but
it can be further improved as there is an observed correlation between the maximum absolute 
brightness of a supernova and the rate at which its brightens and faded. And because a supernova at
maximum brightness has a luminosity comparable to an entire galaxy, they can be seen at great
distance. Exactly such an effect has been observed for several dozen Type Ia high z supernovae
$(z_{max} \leq 0.83)$ by two teams: the supernova cosmology project \cite{ref10} and the high-z
supernova search team \cite{ref11}. The results delivered a major surprise to cosmologists.
None of the usual cosmological models without a cosmological constant were able to explain the
observed luminosity distance curve. The observations of Perlmutter {\it et al.} \cite{ref10} indicate
that the joint probability distribution of $\left(\Omega_{m}, \Omega_{\Lambda}\right)$ is well fitted by
$$ 0.8\Omega_{m}  - 0.6\Omega_{\Lambda} \simeq - 0.2 \pm 0.1.$$ 
The best-fit region strongly favours a positive energy density for the cosmological constant
$\Omega_{\Lambda} > 0$. \\

\section{Angular Diameter Distance}
The angular diameter distance is a measure of how large objects appear to be. As with 
the luminosity distance, it is defined as the distance that an object of known physical
extent appears to be at, under the assumption of Euclidean geometry. \\
 
The angular diameter $d_{A}$ of a light source of proper distance
$d$ is given by
\begin{equation}
\label{eq30} 
d_{A} = d(z)(1+z)^{-1}=d_{L}(1 + z)^{-2}.
\end{equation}
Applying Eq. (27) we obtain
\begin{equation}
\label{eq31} 
H_{0} d_{L} =\frac{A_{0}}{1-A_{0}}\left[\frac{1 - (1
+ z)^{\frac{A_{0}-1}{A_{0}}}}{(1+z)}\right].
\end{equation}
Usually $d_{A}$ has a minimum (or maximum) for some $Z= Z_{m}$. In
Freese {\it et al.} \cite{ref34} model, for example this occurs for $n=2$ and 
$$Z_{m} =  \Big(\frac{3}{2}\Omega_{m}\Big)^{\frac{2}{(3\Omega_{m} - 2)}} - 1 .$$
The maximum $d_{A}$ for our model, one can easily find by setting the value of $\Omega_{m}$. \\

The angular diameter and luminosity distances have similar forms, but have a different
dependence on redshift. As with the luminosity distance, for nearly objects the angular
diameter distance closely matches the physical distance, so that objects appear smaller
as they are put further away. However the angular diameter distance has a much more 
striking behaviour for distant objects. The luminosity distance effect dims the radiation
and the angular diameter distance effect means the light is spread over a large angular area.
This is so-called surface brightness dimming is therefore a particularly strong function
of redshift. \\

\section{Look Back Time}
The time in the past at which the light we now receive from a distant object was emitted is
called the look back time. How {\it long ago} the light was emitted (the look back time)
depends on the dynamics of the universe. \\

The radiation travel time (or look-back time) $t - t_{0}$ for
photon emitted by a source at instant $t$ and received at $t_{0}$
is given by
\begin{equation}
\label{eq32} t - t_{0} = \int^{a_{0}}_{a} \frac{da}{\dot{a}} \; ,
\end{equation}
Equation (11) can be rewritten as
\begin{equation}
\label{eq33} a = B_{0} \: t^{A_{0}}, \;\; B_{0} = constant.
\end{equation}
This follows that
\begin{equation}
\label{eq34} \frac{a_{0}}{a} = 1 + z =
\left(\frac{t_{0}}{t}\right)^{A_{0}},
\end{equation}
The above equation gives
\begin{equation}
\label{eq35} t = t_{0}(1 + z)^{-\frac{1}{A_{0}}}.
\end{equation}
From Eqs. (21) and (35), we obtain
\begin{equation}
\label{eq36} t_{0} - t = A_{0}H_{0}^{-1}\left[1 - (1 +
z)^{-\frac{1}{A_{0}}}\right],
\end{equation}
which is
\begin{equation}
\label{eq37} H_{0} (t_{0} - t) = A_{0}\left[1 - (1 +
z)^{-\frac{1}{A_{0}}}\right].
\end{equation}
For small $z$ one obtain
\begin{equation}
\label{38} H_{0} (t_{0} - t) = z - (1+ \frac{q}{2}) z^{2} + ....
\end{equation}
From Eqs. (35) and (37), we observe that at $z \to \infty$, $H_{0}
t_{0}$ = $A_{0}$ (constant). For $n=2$ and  $A_{0}$ =
$\frac{2}{3}$ gives the well-known ES result
\begin{equation}
\label{eq39} H_{0} (t_{0} - t) =  \frac{2}{3}\left[1 - (1 + z)^{-\frac{3}{2}}\right].
\end{equation}
\section {Conclusions}
In this paper we have analyzed the RW multidimensional stiff fluid
cosmological models with $\Lambda$-term varying with time and of the form $\Lambda =
\beta \left(\frac{\ddot{a}}{a}\right)$. The results for the
cosmological tests are compatible with the present observations.
To solve the age parameter and density parameter one require the
cosmological constant to be positive or equivalently the
deceleration parameter to be negative. This imply an accelerating
universe. The proper distance, the luminosity distance-redshift,
the angular diameter distance-redshift, and look back
time-redshift for the model are presented in the frame work of
multidimensional space time. The model of the Freese {\it et al.}
is retrieved from our model for $n=2$  and the particular choice
of $A_{0}$. Also the ES results are obtained for the case $A_{0} =
\frac{1}{2}$. These tests are found to depend on $\beta$. It is a
general belief among cosmologists that more precise observational
data should be achieved in order to make more definite statements
about the validity of cosmological models (Charlton and
Turner \cite{ref35}). We hope that in the near future, with the new
generation of the telescope, the present
situation could be reversed. \\
\section*{Acknowledgements} The authors
wish to thank Prof. Ravi Kulkarni, Director, Harish-Chandra
Research Institute, Allahabad, India, for providing warm
hospitality and excellent facilities where this work was done.
\newline

\end{document}